\documentclass[fleqn,twoside]{article}

\usepackage{espcrc2}
\usepackage{graphicx}
\usepackage{axodraw}

\newcommand{\be}{\begin{equation}}
\newcommand{\bea}{\begin{eqnarray}}
\newcommand{\ee}{\end{equation}}
\newcommand{\eea}{\end{eqnarray}}
\newcommand{\bpi}{\begin{picture}}
\newcommand{\bce}{\begin{center}}
\newcommand{\epi}{\end{picture}}
\newcommand{\ece}{\end{center}}

\def\chic#1{{\scriptscriptstyle #1}}

\title{BRST-driven cancellations and gauge invariant Green's functions} 

\author{D. Binosi\address[UV]{Departamento de F\'\i sica
Te\'orica and IFIC, Centro Mixto,  
Universidad de Valencia-CSIC,
E-46100, Burjassot, Valencia, Spain} and J. Papavassiliou\addressmark[UV]}

\begin{document}

\begin{abstract}

We  study  a fundamental,  all order cancellation
operating  between graphs of distinct  kinematic nature,
which allows for the construction of gauge-independent
effective  self-energies,  vertices,  and  boxes at arbitrary order.

\end{abstract}

\maketitle

When quantizing  gauge theories in  the continuum 
one must invariably resort to an appropriate gauge-fixing procedure in
order   to  remove  redundant   (non-dynamical)  degrees   of  freedom
originating    from    the   gauge    invariance    of   the    theory.    
Thus,  one  adds   to  the   gauge  invariant
(classical) Lagrangian  ${\cal L}_{\rm I}$ a  gauge-fixing term ${\cal
L}_{\rm GF}$, which allows for the consistent derivation of
Feynman  rules.  At  this point  a new  type of  redundancy  makes its
appearance, this time at the level of the building blocks defining the
perturbative  expansion. In  particular, individual  off-shell Green's
functions  ($n$-point  functions) carry  a  great  deal of  unphysical
information,  which disappears when  physical observables  are formed.
$S$-matrix elements, for example,  are independent of the gauge-fixing
scheme  and  parameters  chosen  to  quantize  the  theory,  they  are
gauge-invariant  (in  the sense  of  current  conservation), they  are
unitary  (in  the sense  of  conservation  of  probability), and  well
behaved at high energies.  On  the other hand Green's functions depend
explicitly (and generally non-trivially) on the gauge-fixing parameter
entering  in the  definition of  ${\cal L}_{\rm  GF}$, they  grow much
faster than  physical amplitudes at  high energies 
and   display   unphysical
thresholds.   Last but  not  least,  in the  context  of the  standard
path-integral  quantization  by  means  of the  Faddeev-Popov  Ansatz,
Green's functions satisfy complicated Slavnov-Taylor identities (STIs)
\cite{Slavnov:1972fg}  involving ghost  fields, instead  of  the usual
Ward  identities generally associated  with the  original gauge
invariance.

The  above observations imply  that in  going from  unphysical Green's
functions to physical  amplitudes, subtle field theoretical mechanisms
are at work, implementing vast cancellations among the various Green's
functions.
Interestingly enough,  these cancellations may be exploited
in   a   very   particular   way   by   the   Pinch   Technique   (PT)
\cite{Cornwall:1982zr,Cornwall:1989gv}:
a  given physical amplitude is
reorganized  into  sub-amplitudes,   which  have  the  same  kinematic
properties       as       conventional       $n$-point       functions
(self-energies, vertices, boxes) but, in addition, they are endowed with
important   physical   properties \cite{Papavassiliou:1995fq}.   
The basic observation, which essentially defines the PT, is that there
exists   a   fundamental  cancellation,   driven   by  the   underlying
Becchi-Rouet-Stora-Tyutin (BRST)  symmetry \cite{Becchi:1976nq}, which
takes  place  between  sets   of  diagrams  with  different  kinematic
properties,  such  as self-energy,  vertex,  and  box diagrams.   This
cancellations  are  activated  when longitudinal  momenta  circulating
inside vertex and box diagrams, generate (by ``pinching'' out internal
fermion lines) propagator-like terms; the latter are combined with the
conventional self-energy graphs in order  to give rise to the aforementioned 
effective Green's functions. It turns out that 
these rearrangements can be collectively captured at any order through
the judicious use of the STI
satisfied by a special Green's function, which
serves as a common kernel to all higher order self-energy and vertex
diagrams \cite{Binosi:2002ft}.

We  will consider  for  concreteness the  $S$-matrix
element  for the  quark--anti-quark elastic  scattering  process $
q(r_1)\bar q(r_2)\to q(p_1)\bar q(p_2)$ in  QCD. We set $q=r_1-r_2=p_1-p_2$,
with $s=q^2$ the  square of the momentum transfer.
The  longitudinal momenta  responsible  for the triggering of the
aforementioned STI 
stem either from the bare gluon propagators or from the
pinching  part $\Gamma^{{\rm  P}}_{\alpha\mu\nu}$ appearing
in  the  characteristic  decomposition  of the  elementary  tree-level
three-gluon   vertex   $\Gamma^{eab,[0]}_{\alpha\mu\nu}(q,k_1,k_2)$ into
\cite{Cornwall:1982zr}      
\bea
\Gamma^{[0]}_{\alpha\mu\nu}&=&  \Gamma^{{\rm
F}}_{\alpha\mu\nu}+ \Gamma^{{\rm
P}}_{\alpha\mu\nu},        \nonumber\\ 
\Gamma^{{\rm
F}}_{\alpha\mu\nu}&=&(k_1-k_2)_\alpha  g_{\mu\nu}+2q_\nu
g_{\alpha\mu}-2q_\mu g_{\alpha\nu},\nonumber    \\    
\Gamma^{{\rm
P}}_{\alpha\mu\nu}&=&k_{2\nu}g_{\alpha\mu}-
k_{1\mu}g_{\alpha\nu}.
\label{PTDEC}
\eea 
The above  decomposition is  to be  carried out  to ``external''
three-gluon  vertices  only, {\it i.e.},   the  vertices  where the  physical
momentum transfer $q$ is entering \cite{Papavassiliou:2000az}.   In
what follows we will carry out 
the analysis  in the renormalizable  Feynman gauge (RFG);  this choice
eliminates  the longitudinal  momenta from  the bare  propagators, and
allows  us  to focus  our  attention on  the  all-order  study of  the
longitudinal      momenta      originating     from      $\Gamma^{{\rm
P}}_{\alpha\mu\nu}$. We will denote by
$\cal{A}$ the  subset of the graphs  which will receive  the action of
the     longitudinal    momenta     stemming     from    $\Gamma^{{\rm
P}}_{\alpha\mu\nu}(q,k_1,k_2)$. We have that
\be
\bpi(0,90)(-65,-65)
\Text(-36,-5.6)[r]{${\cal A}\ = \ $}
\Text(-47,-40)[l]{$=\ ig^2\bar  u(r_1) 
\frac{\lambda^e}{2} \gamma_{\alpha} v(r_2) 
f^{eab} \Gamma^{{\rm P}\!,\,\alpha\mu\nu}(q,k_1,k_2)\times$}
\Text(-47,-60)[l]{$\times\ {\cal T}^{ab}_{\mu\nu}(k_1,k_2,p_1,p_2),$}
\ArrowLine(-12,-5.6)(-33.6,16)
\ArrowLine(-33.6,-27.2)(-12,-5.6)
\Text(-36,16)[r]{$\scriptstyle{r_1}$}
\Text(-36,-27.2)[r]{$\scriptstyle{r_2}$}
\Photon(-12,-5.6)(0.8,-5.6){1.2}{3}
\Text(-12,-10.4)[l]{$\scriptstyle{\alpha}$}
\Text(-12,-0.8)[l]{$\scriptstyle{e}$}
\PhotonArc(14,-6)(12,0,360){1.2}{17}
\GCirc(14,5.2){4}{0.8}
\Text(14,5.2)[c]{$\scriptstyle{\Delta}$}
\Text(3.2,5.2)[c]{$\scriptstyle{\nu}$}
\Text(26,5.2)[c]{$\scriptstyle{\sigma}$}
\GCirc(14,-18){5}{0.8}
\Text(14,-18)[c]{$\scriptstyle{\Delta}$}
\Text(3.2,-18)[c]{$\scriptstyle{\mu}$}
\Text(26,-18)[c]{$\scriptstyle{\rho}$}
\ArrowLine(52,16)(30.4,-5.6)
\Text(56,16)[l]{$\scriptstyle{p_1}$}
\ArrowLine(30.4,-5.6)(52,-27.2)
\Text(56,-27.2)[l]{$\scriptstyle{p_2}$}
\GCirc(30,-6){8}{0.8}
\Text(30,-6)[c]{${\cal C}_{\rho\sigma}$} 
\GCirc(1.6,-5.6){7}{0.8}
\Text(1.6,-5.6)[c]{$\scriptstyle{\Gamma^{\rm P}}$}

\epi
\ee
where   $\lambda^e$   are   the   Gell-Mann   matrices,   and   ${\cal
T}_{\mu\nu}^{ab}$     is    the     sub-amplitude    $g_{\mu}^{a}(k_1)
g_{\nu}^{b}(k_2)\to\bar q(p_1)q(p_2)$, with the gluons {\it off-shell}
and    the   fermions    {\it on-shell}.
In terms of Green's functions we have
(omitting the spinors)
\be
{\cal T}_{\mu\nu}^{ab} = 
{\cal C}^{ab}_{\rho\sigma}(k_1,k_2,p_1,p_2)
\Delta^{\rho}_{\mu}(k_1)\Delta^{\sigma}_{\nu}(k_2).
\label{sgf}
\ee
Clearly,  there  is  an  equal contribution  where  the  $\Gamma^{{\rm
P}}$ is situated on the right hand-side of ${\cal T}$.

\begin{figure}[!t]
\bce
\includegraphics[width=7.5cm]{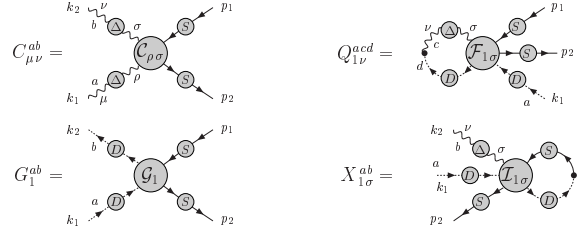}
\vspace{-1.5cm}
\ece
\caption{\label{fig:2} Diagrammatic representation of the
Green's function appearing in the STI of Eq.(\ref{BasSTI}).}
\vspace{-0.4cm}
\end{figure}

Let us focus on the STI satisfied by the amplitude 
${\cal T}_{\mu\nu}^{ab}$.
This STI reads
\bea
k_1^\mu C_{\mu\nu}^{ab}\!\!+\!k_{2\nu}G_1^{ab}\!\!-\!igf^{bcd}
Q_{1\nu}^{acd}
\!\!-\!gX_{1\nu}^{ab}
\!+\!g\bar X_{1\nu}^{ab}\!=\!0,\!\!\! 
\label{BasSTI}
\eea  
where the Green's function appearing in it have the diagrammatic definition
showed in Fig.\ref{fig:2}. The  terms
$X_{1\nu}$ and   $\bar  X_{1\nu}$ 
vanish  on-shell, since  they are  missing one
fermion  propagator. 
Thus,  we
arrive at the on-shell STI for ${\cal T}_{\mu\nu}^{ab}$
\bea
&&\mbox{}\hspace{-0.7cm}k_1^\mu {\cal T}_{\mu\nu}^{ab}=
{\cal S}^{ab}_{1\nu}, \nonumber \\
&&\mbox{}\hspace{-0.7cm}
{\cal S}^{ab}_{1\nu}=
igf^{bcd}{\cal
Q}_{1\nu}^{acd}(k_1,k_2,p_1,p_2)D(k_1)
\nonumber \\
&&\mbox{}\hspace{-0.15cm}
-
k_{2\nu}{\cal
G}_1^{ab}(k_1,k_2,p_1,p_2)D(k_1)D(k_2), 
\label{onshSTI}
\eea
where we have defined 
$Q_{1\nu}^{acd}={\cal Q}_{1\nu}^{acd}
D(k_1)S(p_1)S(p_2)$.

\begin{figure}[!t]
\bce
\includegraphics[width=5.5cm]{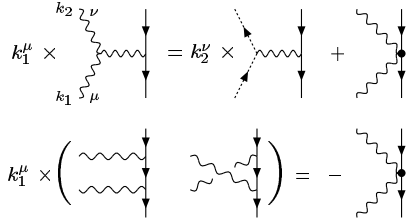}
\vspace{-1.3cm}
\ece
\caption{\label{fig:4} Diagrammatic representation of the tree-level
inequality of Eq.(\ref{INEQ})}
\vspace{-0.4cm}
\end{figure}

In   perturbation   theory  both   ${\cal   T}^{ab}_{\mu\nu}$  and   ${\cal
S}^{ab}_{1\nu}$ are given by  Feynman diagrams, which can be separated
into  distinct classes,  depending on  their kinematic  dependence and
their geometrical properties.  Graphs which do not contain information
about  the  kinematical  details   of  the  incoming  test-quarks  are
self-energy graphs,  whereas those which  display a dependence  on the
test quarks are vertex graphs. The  former depend only on the variable
$s$,  whereas the latter  on both  $s$ and  the mass  $m$ of  the test
quarks; equivalently,  we  will  refer to  them  as $s$-channel  or
$t$-channel  graphs,   respectively.   In  addition   to  the  $s$-$t$
decomposition,  Feynman diagrams  can be  separated  into one-particle
irreducible (1PI) and one-particle  reducible (1PR) ones.  The crucial
point is  that the action of  the momenta $k_1^\mu$  or $k_2^\nu$ on
${\cal T}^{ab}_{\mu\nu}$  does {\it not} respect, in  general, the original
$s$-$t$ and 1PI-1PR separation furnished by the Feynman diagrams.
In other
words, even though Eq.(\ref{onshSTI}) holds for the 
entire amplitude, 
it is not true for the individual sub-amplitudes, 
{\it i.e.},
\be
k_1^\mu \left[{\cal T}^{ab}_{\mu\nu}\right]_{x,{\rm \chic{Y}}} \neq  
\left[{\cal S}^{ab}_{1\nu}\right]_{x, {\rm \chic{Y}}},\quad x=s,t; \,\,\,
{\rm Y=I,R},
\label{INEQ}
\ee
where I (respectively R) indicates the one-particle {\it irreducible}
(respectively {\it reducible}) parts of the amplitude involved.
Evidently,   whereas   the   characterization   of   graphs   as
propagator- and  vertex-like is  unambiguous  in  the absence  of
longitudinal momenta  ({\it e.g.}, in a scalar theory),  their presence tends
to mix propagator- and  vertex-like graphs.  Similarly, 1PR graphs are
effectively converted into 1PI ones (the opposite cannot happen).  The
reason  for  the  inequality   of  Eq.(\ref{INEQ})  are  precisely  the
propagator-like  terms, such  as  those encountered  in  the one-  and
two-loop calculations; they have the characteristic feature that, when
depicted  by means  of Feynman  diagrams contain  unphysical vertices,
{\it   i.e.},  vertices   which   do  not   exist   in  the   original
Lagrangian (Fig.\ref{fig:4}).  All such diagrams  cancel {\it
diagrammatically} against  
each other.  
In particular then,  after the PT cancellations have  been enforced, we
find that the $t$-channel irreducible part satisfies the identity              
\be 
\left[k_1^\mu  {\cal
T}^{ab}_{\mu\nu}\right]_{t,{\rm \chic{I}}}^{\chic{\rm  PT}}  \equiv
\left[{\cal   S}^{ab}_{1\nu}\right]_{t,{\rm \chic{I}}}.
\label{EQPT}
\ee 

The non-trivial step for generalizing the PT to all orders is then the
following: Instead  of going through the arduous  task of manipulating
the  left  hand-side of  Eq.(\ref{EQPT})  in  order  to determine  the
pinching parts and explicitly enforce their cancellation, use directly
the right-hand  side, which already contains the  answer!  Indeed, the
right-hand side involves  only conventional (ghost) Green's functions,
expressed  in terms  of normal  Feynman  rules, with  no reference  to
unphysical  vertices.   Thus,  its  separation  into  propagator-  and
vertex-like  graphs  can  be  carried  out  unambiguously,  since  all
possibility for mixing has been eliminated.

After these  observations, we  proceed to the  PT construction  to all
orders. 
Once  the effective Green's functions have  been derived, they
will be  compared to the  corresponding Green's functions  obtained in
the context of  the Background Field Method Feynman  gauge (BFMFG), in
order  to establish  whether the  known one-  and
two-loop correspondence  persists  to all
orders; as we will see, this is indeed the case (for an extended list
of related references see \cite{Binosi:2002ez}). 

To  begin with,  it is  immediate  to recognize  that in  the RFG  box
diagrams of arbitrary order $n$,  to be denoted by $B^{[n]}$, coincide
with the PT boxes ${\widehat B}^{[n]}$, since all three-gluon vertices
are  ``internal'',  {\it  i.e.},   they  do  not  provide  longitudinal
momenta. Thus, they coincide  with the BFMFG boxes, $\tilde{B}^{[n]}$,
{\it  i.e.}, $  {\widehat B}^{[n]}  = B^{[n]}  =  \tilde{B}^{[n]}$ for
every~$n$.

\begin{figure}[!t]
\bce
\includegraphics[width=7.5cm]{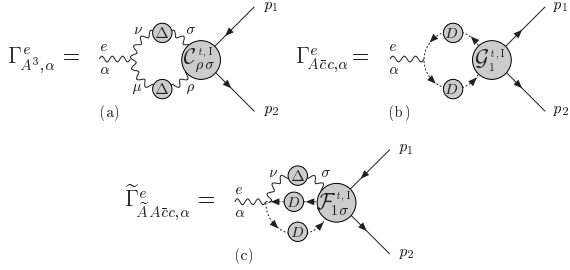}
\vspace{-1.5cm}
\ece
\caption{\label{fig:1} The Green's functions identified in the
construction of the all order PT vertex
$\widehat\Gamma^{e}_\alpha$. 
(b) and (c) receive a contribution from
similar terms with the ghost arrows reversed (not shown).}
\vspace{-0.4cm}
\end{figure}

We  then  continue with  the  construction  of  the {\it  one-particle
irreducible}     PT    gluon-quark--anti-quark     vertex    $\widehat
\Gamma^e_\alpha$.  We start from  the corresponding vertex in the RFG,
to be  denoted by  $\Gamma^e_\alpha$, and focus  only on the  class of
vertex diagrams containing an  {\it external} bare three-gluon vertex;
we   will   denote    this   subset   by   $\Gamma^{e}_{\!A^3,\alpha}$
[Fig.\ref{fig:1}(a)].   All  other  types  of graphs  contributing  to
$\Gamma^e_\alpha$ are inert  as far as the PT  procedure is concerned,
because     they      do     not     furnish      pinching     momenta
\cite{Papavassiliou:2000az}.  The next step is to carry out the vertex
decomposition of  Eq.(\ref{PTDEC}) to the  external three-gluon vertex
$\Gamma^{eab,[0]}_{\alpha\mu\nu}$             appearing             in
$\Gamma^{e}_{\!A^3,\alpha}$.    This  will   result  in   the  obvious
separation       $\Gamma^{e}_{\!A^3,\alpha}       =       \Gamma^{{\rm
F}\!,\,e}_{\!A^3,\alpha} +  \Gamma^{{\rm P}\!,\,e}_{\!A^3,\alpha}$. The
part $\Gamma^{{\rm  F}\!,\,e}_{\!A^3,\alpha}$ is also  inert, and will
be left untouched.  Thus, the  only quantity to be further manipulated
is $\Gamma^{{\rm P}\!,\,e}_{\!A^3,\alpha}$;  it reads 
\be 
\Gamma^{{\rm
P\!,\,}e}_{\!A^3,\alpha}\!\!=\!\! gf^{eba}\int \left[(k-q)^\mu
g^\nu_\alpha+k^\nu g^\mu_\alpha\right] \left[{\cal
T}_{\mu\nu}^{ab}\right]_{t,{\rm I}},   
\ee   
where    $\int   \equiv
\mu^{2\varepsilon}  \int  d^dk/(2\pi)^d$,  with  $d=4-2\varepsilon$,  and
$\mu$ is the 't Hooft mass.  Following the discussion presented above,
the  pinching   action  amounts  to  the   replacement  $(-k+q)^\mu  [{\cal
T}_{\mu\nu}^{ab}]_{t,{\rm       I}}       \to       [(-k+q)^\mu       {\cal
T}_{\mu\nu}^{ab}]_{t,{\rm          I}}          =          \left[{\cal
S}_{1\nu}^{ab}(-k+q,k)\right]_{t,{\rm  I}}$ and  similarly  for the
term coming from the momentum 
$k^\nu$,    or,    equivalently,   
\be    
\Gamma^{{\rm
P\!,\,}e}_{\!A^3,\alpha}\rightarrow  gf^{eba}\! \int \bigg( [{\cal
S}_{2\alpha}^{ab}]_{t,{\rm  I}}  -  [{\cal  S}_{1\alpha}^{ab}]_{t,{\rm
I}}\bigg) \!,
\label{PTvertex}
\ee
being ${\cal
S}_{2\alpha}^{ab}$ the Bose symmetric of the ${\cal
S}_{1\alpha}^{ab}$ term.

At   this  point  the   construction  of   the  effective   PT  vertex
$\widehat\Gamma^{e}_{\alpha}$ has been  completed.  The next important
point is to study the connection between $\widehat\Gamma^{e}_{\alpha}$
and            the           gluon--quark--anti-quark           vertex
$\widetilde\Gamma^{e}_{\alpha}$  in  the BFMFG.   To  begin with,  all
``inert'' terms contained  in the original $\Gamma^{e}_{\alpha}$ carry
over  to the same  sub-groups of  graphs obtained  in the  BFMFG; most
notably, the $\Gamma^{{\rm  F}\!,\,e}_{\!A^3,\alpha}$ is precisely the
$\Gamma^{e}_{\!\widetilde{A}A^2,\alpha}$            part            of
$\widetilde\Gamma^{e}_{\alpha}$,   where    $\widetilde{A}$   is   the
background   gluon.   The  only   exception  are   the  ghost-diagrams
contributing to $\Gamma^{e}_{\alpha}$ [Fig.\ref{fig:1}(b)]; the latter
do {\it  not} coincide with  the corresponding ghost  contributions in
the BFMFG.
The  important step is  to recognize  that the  BFMFG ghost  sector is
provided  precisely  by combining  the  RFG  ghosts  with the
right-hand side  of 
Eq.(\ref{EQPT}).  Specifically,  one arrives  at both  the {\it  symmetric}
vertex $\widetilde\Gamma_{\widetilde{A}\bar  c c}^{e}$, characteristic
of       the      BFMFG,      as       well      as       at    the
background-gluon--gluon--ghost--anti-ghost                      vertex
$\widetilde\Gamma_{\!\widetilde{A} A\bar  c c}^{e}$, which  is totally
absent in the conventional formalism [Fig.\ref{fig:1}(c)].  Indeed,
using Eq.(\ref{PTvertex}),  we
find 
\bea
&&\mbox{}\hspace{-0.7cm}
\widetilde\Gamma^{e}_{\!\widetilde A\bar c
c,\alpha}\!(q)\!\equiv\!
\Gamma^{e}_{\bar c
c,\alpha}(q)\!-\!gf^{eba}\!\!\!\int \!\!\!\left\{\!k_\alpha\!\left[ 
{\cal G}_1^{ab}(-k+q,k)\!\right]_{t,{\rm
I}}\right.\nonumber \\
&&
\mbox{}\hspace{-0.7cm}
\left.-(k-q)_\alpha\left[
{\cal G}_2^{ab}(-k+q,k)\right]_{t,{\rm
I}}\right\}D(-k+q)D(k), \nonumber \\
&&\mbox{}\hspace{-0.7cm}
\widetilde\Gamma^{e}_{\!\widetilde AA\bar c
c,\alpha}\!(q)\!\equiv\!
ig^2\!f^{eba}\!\!\!\!\int \!\!\!\left\{\!f^{acd}\!
\left[\!
{\cal Q}_{2\alpha}^{cdb}(-k+q,k)\!\right]_{t,{\rm
I}}\!\!D(k)\right. \nonumber \\
&&\mbox{}\hspace{-0.7cm}
\left.  
+f^{bcd}
\left[
{\cal Q}_{1\alpha}^{acd}(-k+q,k)\right]_{t,{\rm
I}}D(-k+q)\right\}.
\eea
This   last   step   concludes    the   proof   that   the   equality
$\widehat\Gamma^{e}_\alpha\equiv\widetilde\Gamma^{e}_\alpha$    between
the PT and  BFMFG vertex holds true to all  orders.  We emphasize that
the  sole ingredient in  the above  construction has  been the  STI of
Eq.\ref({onshSTI});  in particular,  at  no point  have  we employed  {\it
a priori} the  BFM formalism.  Instead,  the special BFM  ghost sector
has  arisen {\it  dynamically},  once the  PT  rearrangement has  taken
place.

The final step  is to construct the (all  orders) PT gluon self-energy
$\widehat\Pi^{ab}_{\mu\nu}$.   Notice  that at  this  point one  would
expect  that  it  too  coincides  with  the  BFG  gluon  self-energy
$\widetilde\Pi^{ab}_{\mu\nu}$,  since both  the boxes  as well  as the
vertex do coincide with the corresponding quantities in BFG, and the
$S$-matrix is  unique. In fact this has been shown to be the case 
both through an inductive proof as well as by a direct construction
\cite{Binosi:2002ft}. 

In conclusion, we have shown that the use of the underlying BRST symmetry
allows (trough the PT algorithm) for the construction of gauge
independent and gauge invariant Green's functions in QCD.
It would be interesting to further explore the physical meaning of 
the $n$-point functions obtained \cite{Bernabeu:2002nw}, 
and establish possible connections with related 
formalisms. 

{\bf Acknowledgments:}
This work has been supported by the CICYT Grants
AEN-99/0692 and FPA-2002-00612. ~J.P. thanks the organizers of QCD~03
for their hospitality and for 
providing a very pleasant and stimulating atmosphere.

\end{document}